 \documentclass[12pt,preprint]{aastex}

\slugcomment{Submitted to ApJ}

\shorttitle{The Parallax of RX J185635-3754 Revisited}
\shortauthors{Walter et al.}
\newcommand{\ud}[2]{\mbox{$^{+ #1}_{- #2}$}}

\begin{document}

\title{Revisiting the Parallax of the Isolated Neutron Star RX J185635-3754 
Using HST/ACS Imaging}

\author{F.M. Walter }
\affil{Department of Physics and Astronomy, Stony Brook University,
    Stony Brook, NY 11794-3800}

\author{T. Eisenbei\ss }
\affil{Astrophysikalisches Institut und Universit\"ats-Sternwarte, Friedrich-Schiller-Universit\"at, Jena, DE}

\author{J.M. Lattimer \& B. Kim }
\affil{Department of Physics and Astronomy, Stony Brook University,
    Stony Brook, NY 11794-3800}

\author{V. Hambaryan \& R. Neuh\"auser }
\affil{Astrophysikalisches Institut und Universit\"ats-Sternwarte, Friedrich-Schiller-Universit\"at, Jena, DE}

\begin{abstract}

We have redetermined the parallax and proper motion of the nearby
isolated neutron star RX~J185635-3754. We used eight observations with the high
resolution camera of the HST/ACS taken from 2002 through 2004. 
We performed the astrometric fitting using five independent methods, all
of which yielded consistent results.
The mean estimate of the distance is 123\ud{11}{15}~pc ($1~\sigma$), 
in good agreement with our earlier published determination.

\end{abstract}


\section{Introduction}

Distance is one of the most fundamental attributes of an astrophysical
source. The conversion of most observable quantities (e.g., flux, angular
diameter, proper motion) to invariate physical units
(e.g., luminosity, linear diameter, space velocity) depends
the distance to some power.
The distance can be determined in a model-independent manner using the
trigonometric parallax $\Pi$.
This geometrical measure of the distance 
is one of the most demanding measurements in all of astrophysics because the
angles involved are so small.

Here we revisit the distance to the isolated neutron star RX~J185635-3754.
This object is one of the so-called ``Magnificent 7'' \citep{Ha07},
a collection of thermally-emitting isolated compact objects. 
These objects are of interest because,
in principle, we can estimate their angular radii from the observed flux
and atmospheric models. Measurement of the radius is a necessary
step in determining the mass-radius relation for neutron stars (if indeed they
are neutron stars and not some more exotic object) and
constraining the equation of state of matter at nuclear densities \citep{LP01}.

Because these isolated stars neither accrete from a binary companion
nor have spectra dominated by non-thermal emission, 
they afford our best opportunity to examine the surfaces of neutron stars. 
If the mass can also be estimated, say, from a gravitational
redshift, then within the uncertainties one can tightly constrain the nuclear
equation of state.
Even without a mass constraint, an accurate radius measurement
(to within 10\%) is sufficient to constrain significant, and to date
uncertain, nuclear matter properties such as the density dependence of
the nuclear symmetry energy.

The population of isolated neutron stars consists of objects that
are still sufficiently young ($\la 10^6$ years) that the thermal emission from
their surfaces is detectable in soft X-rays. Older objects both cool (shifting
the emission into the largely unobservable EUV) and fade. In the optical
these young objects have absolute magnitudes of about 20.
By an age of a few million years, an X-ray-bright isolated neutron star with
a heavy element surface
will have cooled from 10$^6$K to 10$^5$K, with a corresponding
10 magnitude decrease in brightness \citep{P04}.

The isolated, radio-quiet, thermally-emitting soft X-ray source
RX~J185635-3754 was
discovered by \citet{WWN96}. The optical counterpart,
identified by \citet{WM97}, is a blue point source with V$\sim$25.7.
Unlike the other members of the
Magnificent 7, there is no evidence for any absorption features in the X-ray
spectrum. There must be surface inhomogeneities, however, as the object has
been seen to rotate with a 7 second period \citep{TM07}.
The magnetic field is inferred to be of order 10$^{13}$G \citep{vKK08}, based on
a very uncertain measurement of the spindown.
While the X-ray spectrum is indistinguishable from black body with
kT$_\infty$=63~eV
\citep{B03}, and the UV/optical spectral energy is consistent with
the Rayleigh-Jeans tail of a hot (T$_\infty>6.5$~eV) black body \citep{P01},
the two spectra do not 
match. The long wavelength flux is about a factor of 7 larger than an
extrapolation of the X-ray blackbody into the optical.
Aside from the lack of any features in the X-ray spectrum, these properties are
similar to those of the other members of this small class of
isolated neutron stars.

A number of lines of reasoning point to RX~J185635-3754's
being a nearby compact object. 
Perhaps the most compelling is that the measured interstellar extinction
is small, yet the line of sight passes near the Corona Australis star forming
region and its associated molecular clouds, which are at a distance of about
120--140~pc \citep{MR81, NF08}.
Because this distance is amenable to direct measurement, and
because of the fundamental importance of constraining the equation
of state at nuclear densities, we set out to measure the parallax of this
object.

\citet{W01} published a parallax of 16.5$\pm$2.3 milli-arcsec (mas), which
corersponds to a distance of 61~pc,
based on 3 HST/WFPC2 images.
This was later found to be erroneous.
\citet{KvKA02} published a parallax of 7$\pm$2 mas (140~pc).
After incorporating a fourth WFPC2 image,
\citet{WL02} reported a revised parallax of 8.5$\pm$0.9 mas
(117$\pm$12~pc). 
Based on models of the interstellar extinction, \citet{P07} find likely
distances between 125 and 135~pc.

The radius of the neutron star is the product of the
model-dependent angular radius and the distance.
The simplest models, blackbody fits to the X-ray spectrum, yield
unrealistically small radii of a few km, and ignore information contained in
the longer wavelength spectra. 
The isolated neutron stars have
optical excesses relative to their X-ray fluxes
(e.g., Pons et al.\ 2001, Haberl 2007);
models that have been considered to explain the spectrum include
two-component black bodies,
a condensed magnetized heavy element surface \citep{PMP05}, and
optically thin partially-ionized hydrogen atop a condensed surface \citep{Ho07}.
\citet{WL02} fit a simple two-blackbody model.
They showed that at 117~pc, the radiation radius
R$_\infty$ is $<$26 km (it depends on the ill-defined temperature of the
cooler component), with a formal best-fit radius of 16.4$\pm$1.7~km, where
the uncertainty reflects that of the distance. \cite{Tr04} find a conservative
lower limit to the radius of 16.5~km for a two blackbody fit at the same
distance.
\citet{Ho07} fit their magnetic H-atmosphere plus
condensed surface model, and find R$_\infty$=17~km, assuming a distance of
140~pc (14 km at a distance of 117~pc).

Seeing a need to confirm this measurement, Kaplan et al.\
re-observed the target with the High Resolution Camera (HRC)
of the Advanced Camera for Surveys (ACS) on the Hubble Space Telescope.
The factor of two improvement in plate scale affords the opportunity to
refine the measurement, reduce the uncertainty in the distance, and yield
improved physical parameters for the closest isolated neutron star.
\citet{KvKA07} and \citet{vKK07} reported distances of 167\ud{18}{15}~pc
and 161\ud{18}{14}~pc, respectively, without providing
details.

In this paper we analyze these ACS/HRC images to measure the parallax of
RX~J185635-3754 at a second epoch.

\section{Observations}

The target was observed with the ACS/HRC on eight
occasions over a 20 month span,
beginning in 2002 September (Table~\ref{tbl-times}). 
The HST program ID is 9364; the PI is D.L.~Kaplan. 

There are two requirements for astrometric observations: a large plate scale
because the motions are small, and a large field of view, to maximize the
number of background reference objects.
The HRC channel of the ACS is a 1024$\times$1024 pixel CCD camera with a
nominal pixel size of about 28 milli-arcsec. It was presumably
selected for these observations because of the large plate scale. 
The relatively small field of view and commensurately small number of
potential reference stars is the main limitation on the accuracy
with which the images
can be coaligned, or transformed, to a common reference frame.

At each visit four
observations were taken using the standard 4-point dither pattern with 2.5 pixel
offsets in each coordinate. Each integration was 620 seconds through the 
F475W filter. Images were not CR-SPLIT.

All observations were obtained in Fine Lock
mode. There were no guiding anomalies. The RMS jitter varies between 1.4 and 
3.4~mas. While some of the JIF images show minor elongation of the psf, this
will not affect the positional measurements.

The visits were scheduled $\pm$ 1 month from the extrema of the parallactic 
displacement in right ascension.
At the ecliptic latitude of the target, the position angle of the ellipse
is 83$^\circ$ with an aspect ratio of about 5:1. Consequently the parallax 
is better detectable in residuals in right ascension than in declination.

\subsection{Methodology}

We measure the residual motions of the target differentially, with respect to
the other stars in the images. We assume that there is no bulk motion of the
other stars -- that
is, the directions of the proper motions of the reference stars are 
randomly directed. In the pathological case that all the reference stars
are members of a moving group, with a shared proper motion and identical
parallaxes, then the method of differential astrometry would fail as the shared
motions would be interpreted as detector offsets.
There is no evidence that this is the case. 
Note that a bulk motion will
only affect the derived proper motions. We show later that the reference stars
in this field are at distances $>$10~kpc, so the correction from relative to
absolute parallaxes must be small. 

Because the parallax is small -- a fraction of a pixel in size -- we need to
take great care with the reductions. Therefore, we decided
that each of the investigators would analyze the data independently, using
different fitting techniques. We started from a common
set of processed images.

\subsubsection{Initial Data Processing}

The initial data processing consists of cleaning the images of cosmic rays,
and correcting the pixel positions for distortions in the detector.

Cosmic rays are a problem with all CCD detectors. While largely
cosmetic, occasionally a cosmic ray does hit within the point spread function
of a source, and will affect both the astrometry and the photometry.
We begin with the flat-fielded \_flt.fits images.
We reject all pixels with a data quality flag set to 2048 or greater
(those identified as saturated pixels and cosmic rays)
by setting the data value equal to the 
median of the surrounding 8 pixels.
About 2\% of the pixels in each image are so-affected.
This is mostly done to simplify the data
reduction code, where large data values can draw off a mean or median. But in
the end this is of little consequence.
We save this cosmetically-corrected image as a fits file.

The ACS is a radial bay instrument on the HST, hence there is significant and
asymmetric distortion in the images \citep{AK04}. We use
Anderson's {\it img2xym\_HRC} Fortran code to find the stars in the
cosmetically-corrected images and to correct for the instrumental distortion.
This code determines positions by doing a PSF-fit using the filter-specific
point spread function.
According to \citet{AK04}, the distortion correction corrects to better
than 0.01 pixel in each coordinate for sufficiently bright stars.
The output of this code is a list of raw and corrected X and Y positions in 
the instrumental frame, along with an instrumental magnitude.
The code does not return any estimate of the
uncertainty in the position. We adopt the mean pixel scale of
0.02827~arcsec/pix.

Using the thresholds we selected (HMIN=5, FMIN=150),
we identify 21 stars in the field that are
common to at least five of the visits, in additional to the neutron star.
Thirteen of these stars are recovered
in all visits, and five are seen in seven of the visits.
The others lie close enough to edge of the detector that 
instrumental rotation leaves them outside the field of view on occasion.
Only the fifth visit contains all 21 stars.

\begin{figure}
\plotone{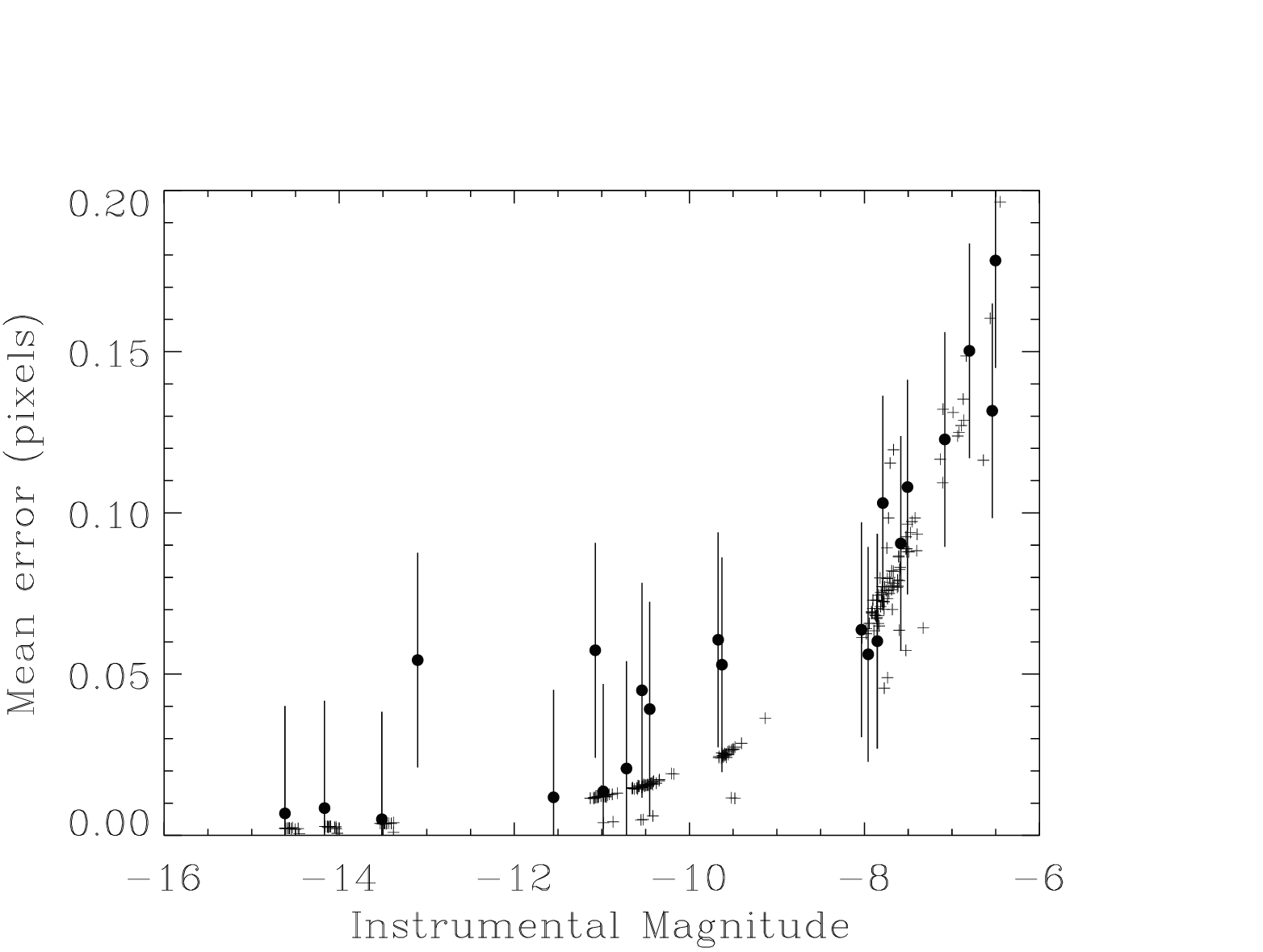}
\caption{The median uncertainty in the accuracy of the transformations as a
function of instrumental magnitude (solid circles).
The uncertainty is computed as the scatter
in the position after transformation at each epoch. There is a single
uncertainty for each star at each epoch, hence 8 points contribute to the
median. 
The small + symbols are the formal uncertainties on the positions, as
measured using the IDL starfinder procedure.
\label{fig-err_imag}}

\end{figure}

We estimate the measurement uncertainties $\sigma$
at each iteration of the various fits.
We take the measurement error on a particular star to be the
standard deviation of the position measurements in the 4 images obtained during
a visit after transforming to a common frame.
As the transformation changes, so do these uncertainties.
Where a positional measurement disagrees by more than one pixel from the
mean of the other three, we flag that measurement as bad and ignore it. Most
such instances are due to cosmic rays affecting the initial PSF fit.
The uncertainties $\sigma$ vary from star to star and for each visit, but 
are identical for the 4 measurements of a star in a particular visit.
We set a minimum $\sigma$=0.01 pix, the accuracy to which \citet{AK04}
claim to be able to measure the positions.
As expected, $\sigma$ increases with instrumental magnitude
(Figure~\ref{fig-err_imag}. In this plot we also show the formal uncertainties
on the positions as extracted using the IDL {\it StarFinder} procedure
\citep{DB00}.
The two agree well.

The initial transformations are accomplished by first shifting the measured
positions within a given visit by the nominal dither, and then 
rotating the X,Y coordinates of the stars to nominal north using the
ORIENTAT keyword in the fits images.
We wrote four independent $\chi^2$ minimization astrometric fitting codes, and
wrote another incorporating the IDL function
MPFITFUN\footnote{MPFITFUN is part of the MPFIT
data fitting package, written by C. Markwardt, and available at
http://www.physics.wisc.edu/\~{}craigm/idl/fitting.html. This is a non-linear
fitting code using $\chi^2$ minimization that should be functionally
equivalent to the technique described herein.}.
Our fitting codes
differ in small details, including how they solve for the parallax ellipse,
how they reject outliers, and how they estimate the measurement uncertainties.
We refer to these models as A1 through
A4 in Table~\ref{tbl-results}. As a test, we also fit the data using a
publicly-available astrometric package, ATPa (model T).
Finally, we generated an independent Bayesian estimate of the parallax
and proper motion (referred to as model B).
The various models yielded indistinguishable results.

\subsubsection{Astrometric Fitting by $\chi^2$ Minimization}

This is a general description of astrometric fitting via $\chi^2$ minimization.
In detail this corresponds to model A1. Differences between the models are
described below.
Our notation is such that $i$ refers to a star, $k$
refers to an epoch, and $j$ refers to one of the four images at that
epoch.  There are a total of $JK=32$ images and $I=21$ field stars.

We register all images by solving for their centers and relative
rotations. The plate scale is not necessarily the same for all images.
We considered the case where the scale varied independently in the X and Y
directions, but found no need for this. We limited our fitting to a scalar
scale change.
The stars have proper motion and a correction for parallax might need
to be included.

The measured positions form the data sets $x_{i,j,k}$ and $y_{i,j,k}$.
If the positions of all field stars appeared on all frames there would
be a total of 672 $(x,y)$ points, but the positions of some stars on
some frames are invalid, either because they fall off the frame, or because they
exceed the 1 pixel error tolerance we adopted.

The function to be minimized is
\begin{equation}
\chi^2=\sum_{i,j\ne0,k\ne0}\left [{(X_{i,j,k}-x_{i,j,k})^2
\over\sigma_{x,i,j,0}^2+\sigma_{x,i,j,k}^2} + {(Y_{i,j,k}-y_{i,j,k})^2\over\sigma_{y,i,j,0}^2+\sigma_{y,i,j,k}^2}\right ]
\end{equation}
where the sum is over the valid indices.  The model
predictions for the positions, relative to a reference position 
$(x_{i,0,0},y_{i,0,0})$, are
\begin{equation}
\begin{array}{ll}
X_{i,j,k}=M_{j,k}[\cos(\theta_{j,k})x_{i,0,0}- \sin(\theta_{j,k})y_{i,0,0}+\Delta x_{j,k}+\mu_{x,i}t_k+P_{x,k}\Pi_i], \\
Y_{i,j,k}=M_{j,k}[\cos(\theta_{j,k})y_{i,0,0}+ \sin(\theta_{j,k})x_{i,0,0}+\Delta y_{j,k}+\mu_{y,i}t_k+P_{y,k}\Pi_i]. 
\end{array}
\end{equation}
The reference position is
taken to be the first frame ($j=0$) of either the first or fifth epochs
($k=0$ or 4) except when
the stellar data for that frame is invalid, in which case the next
frame of the epoch is used. In all cases, we use the notation $0,0$ to refer
to the reference image.  $M$ refers to the plate's relative scale factor, or
magnification, $\theta$ to the plate's relative rotation, $\mu_x$ and
$\mu_y$ to the star's proper motion, $t$ to the difference in epoch
between an image and the reference image, and $\Pi=1/D$ where $D$ is the
stellar distance.  $P_x$ and $P_y$ are the relative parallactic
displacements which depend on the time of year.
If the parallax $\Pi$ is in units of pc$^{-1}$, then
$P_x$ and $P_y$ have units of pixels-pc.  With these definitions
$M_{0,0}=1$, $\Delta x_{0,0}=\Delta
y_{0,0}=\theta_{0,0}=t_0=P_{x,0}=P_{y,0}=0$.  It follows that
$X_{i,0,0}=x_{i,0,0}$ and $Y_{i,0,0}=y_{i,0,0}$.  Initially, it is
assumed $\mu_{x,i}=\mu_{y,i}=\Pi_i=0$ for all $i$.

The measurement errors, $\sigma_{x,i,k}$ and $\sigma_{y,i,k}$,
are the standard deviations among the measured
positions for each star at each epoch, which normally numbers
$J_{i,k}=4$.   
\begin{equation}
\begin{array}{ll}
\sigma_{x,i,k}=\sqrt{\sum_j(X_{i,j,k}- x_{i,j,k})^2/J_{i,k}},\\
\sigma_{y,i,k}=\sqrt{\sum_j(Y_{i,j,k}- y_{i,j,k})^2/J_{i,k}}
\end{array}
\end{equation}
These errors are same for all values of $j$ for each value of $k$. 
As discussed above, we set a floor of 0.01 pix as the 
minimum value for $\sigma$.

The plate registration is done in the standard way by solving the
$4(JK-1)=124$ simultaneous equations for the minimization of $\chi^2$:
\begin{equation}
\begin{array}{llll}
{\partial\chi^2\over\partial\Delta x_{j,k}}=0,\\
{\partial\chi^2\over\partial\Delta y_{j,k}}=0,\\
{\partial\chi^2\over\partial\theta_{j,k}}=0,\\
{\partial\chi^2\over\partial M_{j,k}}=0.
\end{array}
\end{equation}
Note that in these equations, $(j,k)\ne(0,0)$.  

We use a Newton-Raphson 
approach to solving the set of equations $f_i(x_j)=0$.  
Assume an initial guess $x_{0j}$ and make the Taylor expansion to
determine the corrections $dx_{0j}$:
\begin{equation}
f_i(x_{0j})+{\partial f_i\over \partial x_j}dx_{0j}=f_{0i}+A_{ij}dx_{0j}=0.
\end{equation}
Solving this system gives
$dx_{0j}=-(A_{ij})^{-1}f_{0i}.$ The new guess becomes
$x_{1j}=x_{0j}+dx_{0j}$, a new Taylor expansion is made, and a new
correction $dx_{1j}$ is estimated and so forth until convergence is
achieved.

Next, the stellar parameters, their proper motions and distances, are determined
from the $3I=63$ simultaneous equations
\begin{equation}
\begin{array}{lll}
{\partial\chi^2\over\partial \mu_{x,i}}=0,\\
{\partial\chi^2\over\partial \mu_{y,i}}=0,\\
{\partial\chi^2\over\partial \Pi_{i}}=0.
\end{array}
\end{equation}
 Now using the non-zero
values for $\mu_{x,i}, \mu_{y,i}$ and $\Pi_i$, the positional errors and
plate registrations are repeated.  The scheme converges after about 4
iterations.

The target star parameters are established once convergence is
achieved by minimizing
\begin{equation}
\chi^2_{\rm NS}=\sum_{j,k}\left[{(X_{{\rm NS},j,k}-x_{{\rm NS},j,k})^2
\over\sigma_{x,{\rm NS},j,0}^2+\sigma_{x,{\rm NS},j,k}^2} +
{(Y_{{\rm NS},j,k}-y_{{\rm NS},j,k})^2
\over\sigma_{y,{\rm NS},j,0}^2+\sigma_{y,{\rm NS},j,k}^2}\right],\end{equation}
utilizing the above expressions for $i=NS$ and solving the equations
for $i=$NS, where NS is the index corresponding to the neutron star.  

\subsubsection{Transform Uncertainties}

The transformations are based on between 18 and 21 stars in each visit.
The median residual error in the transformation
(the difference between the modeled
stellar positions and the reference positions) is 0.05 pixels; the median
measurement error is 0.07 pixels. 
In principle 
the accuracy of the transform depends on the instrumental magnitude of the
stars used. Examining the spread of errors, it is clear that the 12 stars with
instrumental magnitudes $<-9$ exhibited less scatter in the residuals. Including
only these bright stars gives a median residual error 0.04 pixels with a 
median measurement error of 0.02 pixels. However, no systematic uncertainties
are introduced by using all the stars in the field, and neither the
results nor their uncertainties depend on the choice of reference stars.

\subsubsection{Determination of Parallactic Ellipse}  

We determine the parallactic ellipse using standard formulae
(e.g., in the Astronomical Almanac), for the nominal coordinates
of the target,
$\alpha=18^{\rm h}56^{\rm m}35.11^{\rm s}$ and $\delta=-37^\circ 54' 30.5''$.

\subsubsection{Differences in the models}

In model A2 we use only the nine bright reference stars that appeared in all 32
images to determine the image transformations.
This model uses an iterative approach wherein the plate
transformation is fit with all stellar motions fixed at zero. The proper
motion is then estimated as a linear trend in the residuals.
This is fit, subtracted,
and the plate is re-transformed. This process is repeated until the solution
converges. The neutron star is not included in the fit. This approach
implicitly
assumes that all parallaxes are zero. In the second step the 
parallax and proper motion are fit to the transformed positions of the neutron
star. 
Finally, this process is repeated using all the stars in the image.
No significant parallaxes were found, validating the first assumption. 


In model A3 we estimate the individual uncertainties empirically 
in the $\_$flt images using the
IDL xstarfinder procedure \citep{DB00}.
An empirically determined PSF is fitted to each star in each image, resulting
in astrometric uncertainties ranging from $\sim$0.01 pixels for bright objects
to $\sim$0.09 pixels for the faintest objects, including the target.
We use these uncertainties along with the distortion-corrected positions
from the {\it img2xym\_HRC} output.
The first image in the first visit is used as a reference frame. We fit 
the 19 stars visible in the first image (including the neutron star) and
performs a $\chi^2$ minimization as described above, fitting the proper motions
and parallaxes simultaneously. 

Model A4 is based on MPFIT.
This model can be run using either some
(we selected the brightest 12 stars; those
with instrumental magnitudes $< -9$)
or all the stars in the field for the determining the transformations.
An iterative approach is used, similar to that used in model A2, except that
the parallaxes and proper motions are fit simultaneously. Parallaxes can be
forced to be zero.


\subsubsection{Bayesian Error Estimation (Model B)}

We transformed the positions of all stars in the images to the reference frame
(the first image of the first visit) allowing for 5 free parameters
(translations and scale factors in both in X and Y and orientation). We rejected
stars which failed to transform (including the target), leaving a subsample
of 12 stars to define the reference frame. We estimated the measurement
uncertainties by fitting a straight line to the mean errors of these 12 stars
as a function of instrumental magnitude. This differs from the scheme used
in the A models.

Thus, having positional measurements and corresponding uncertainties
of the target in the common reference of frame,
we computed posterior probabilities of the
parallax and proper motions of the target
(all other parameters are treated as nuisance parameters, which are
 not of immediate interest but which must be accounted for in the analysis 
of those parameters which are of interest). 

The  expected geocentric  right  ascension  and
declination  $\alpha_{GCRS}(t),\delta_{GCRS}(t)$ of the target at a
given epoch $t$ can be expressed in the form:

\begin{equation} \left
\{ 
\begin{array}{ll}
\label{eqs}
   \alpha_{GCRS}(t)  =  \alpha_{BCRS}(t_0) + \Delta\alpha_1(t,t_0,\alpha,\delta,\
mu_{\alpha},\mu_{\delta},\Pi,v_r) + \Delta\alpha_2(t,\alpha,\delta,\Pi) \\
   \delta_{GCRS}(t)  =  \delta_{BCRS}(t_0) + \Delta\delta_1(t,t_0,\alpha,\delta,\
mu_{\alpha},\mu_{\delta},\Pi,v_r) + \Delta\delta_2(t,\alpha,\delta,\Pi) \\
\end{array}
\right. \end{equation}

\noindent
where  $\alpha_{BCRS}(t_0),  \delta_{BCRS}(t_0)$  are  the
barycentric coordinates at  a reference time  $t_0$, the pairs
($\Delta\alpha_1,\Delta\delta_1$) and  ($\Delta\alpha_2,\Delta\delta_2$)
are the corresponding  displacements of the star's position in the 
right  ascension  and declination owing to the spatial motion and 
parallactic motion of the Earth. 
$(\mu_{\alpha},\mu_{\delta})$  are  the right  ascension  and
declination  components  of the   proper motion,  $\Pi$ is the
annual parallax and $v_r$ is radial velocity.

In principle, those expected positions and actual measured ones in different
epochs
may differ by some uncertainty $\epsilon_i$ owing to the measurements errors
plus any real signal in the data that cannot be explained by the model:

\begin{equation} 
\left
\{ 
\begin{array}{ll}
\label{eqs2}
   \alpha_{GCRS}({\rm expected})=\alpha_{GCRS}({\rm measured}) +\epsilon_{\alpha}(t) \\
   \delta_{GCRS}({\rm expected})=\delta_{GCRS}({\rm measured}) +\epsilon_{\delta}(t) 
\end{array}
\right. 
\end{equation}

In the absence of detailed knowledge of the effective noise distribution, 
the most conservative choice of the distribution of positional differences
($\epsilon_i$) 
is a Gaussian and therefore for the likelihood we may write:

\begin{equation} 
\label{eqs3}
   p(D|M(\theta),I)=\prod_{i=1}^{n_{obs}}\frac{1}{2\pi\epsilon_i}\exp\left[-\sum_{i=1}^{n_{obs
}}(d_i-d_i({\rm model})))^2/(2\epsilon_i^2)\right],
\end{equation}

\noindent where by $D(d_i, i=1,n_{obs})$ we denote either the right ascension or
the declination of the target at a given epoch.

Using Bayes theorem we may estimate posterior probabilities of parameters:

\begin{equation} 
\label{eqs4}
   p(M(\theta)|D,I)=\frac{p(D|M(\theta),I)p(M(\theta),I)}{p(D|I)},
\end{equation}

\noindent where M($\theta$) is a model with parameters 
${\boldmath \theta}(\alpha_{BCRS}(t_0),\delta_{BCRS}(t_0),\Pi,\mu_{\alpha},\mu_{\delta},V_r)$, 
and $p(M(\theta),I)$ the prior probabilities of the considered parameters.
$p(D|I)$ serves as the normalization.

We used the flat priors for the proper motions and the uninteresting parameters.
For the parallax we used both so-called flat priors and the Jefferys priors.
The flat prior is a uniform probability distribution within some reasonable
interval.
In the Jeffreys prior, formulated as
\begin{equation} 
\label{eqs5}
p(\Pi|I) = 1/(\Pi*Log[\Pi_{\rm hi}/\Pi_{\rm lo}]) ,
\end{equation}
the probability density is uniform in the logarithm of the
parallax ($\Pi_{\rm hi}$ and $\Pi_{\rm lo}$ are the upper and lower bounds of
the interval considered for the parallax).
Use of this prior is motivated the invariance 
of conclusions with respect to scale changes in time
\citep{J98}.
This prior is also form-invariant with respect to
reparameterization in terms of distance. That is,
an investigator working in terms of distance and using a 1/distance prior
will reach the same conclusions as an investigator working in terms of
Parallax and using a 1/Parallax prior.
This would not be true for a prior of another form, for example, a uniform
prior.
To the extent that parameterization in terms of Distance and Parallax are
both equally ``natural'', this form of invariance is desirable.
Of course, if the prior range of a parameter is small when expressed as a
fraction of the center value, then the conclusions will not differ
significantly from those arrived at by assuming a uniform prior.

Since the measurement errors play a crucial role in the estimation 
of the unknown parameters, we considered three different cases: 

\noindent i) positional uncertainties are known for each epoch (from scaling the
uncertainties as a function of instrumental magnitude, as described above).\\
ii) positional uncertainties are unknown but are the same at all
epochs ($\epsilon_i=const\times\sigma$). \\ 
iii) positional uncertainties are unknown but are different for
different epochs ($\epsilon_i=const\times\sigma_i$).

\subsection{Tests}

\subsubsection{Plate Scale}

We checked the plate scale in the reference image
by using 5 stars that are in common with
the 2MASS catalog. The mean plate scale is 28.38$\pm$0.53~mas/pixel, where
the uncertainty is the standard deviation of the 10 independent measurements.
There are 4 stars common with the NOMAD-1 catalog \citep{Za05}. The mean
plate scale is 29.55$\pm$1.3~mas/pix.
These are both consistent with the mean plate scale from \citet{AK04}.

\subsubsection{Parallax Check}
To test our sensitivity to the small expected parallax, we ran a series of
tests wherein we
replaced the target with a test star of known
parallax and proper motion after transforming the frames.
We selected a large proper motion similar to that expected from the target.
We applied a Gaussian-distributed error 
to the position of the test star, and then fit the data in an attempt to
recover the known motions. Results are shown in Table~\ref{tbl-sims}.

\subsubsection{Proper Motion Check}
The 8 visits to the field occurred at 4 distinct times of the year, within
tolerances of 10 days.
On a given date, the parallactic factor is the same for all the stars, so we
can assume that any difference in position is attributable solely to
proper motion. We determined a mean proper motion for each star from the average
of these four measurements for each star in the field. 
Within the uncertainties, the proper
motion of the target is fully consistent with those from the
global minimizations.

The mean proper motions for the 21 field stars are $\mu_\alpha = -2.4 \pm 4.7$,
$\mu_\delta = 0.4 \pm 3.0$ milli-arcsec/yr. There is no evidence for systemic
motions of the field stars.

\subsubsection{Tolerance to Rotation and Magnification}

The full fit solves for four global parameters
(zero-point translations, rotation, and plate scale).
The largest change in plate scale relative to the reference image is 
6$\times$10$^{-5}$; the largest change in orientation from that of the
reference image is 0.0055 degrees (20~arcsec). Neither of these are significant.
We also ran the fitting routines for the cases where the rotation or the 
scale factor were assumed not to change. This did not affect the results
significantly.

We also solved for the case where the parallax of the reference stars is
fixed at zero. This also had no significant effect on the results.

\subsubsection{Publicly Available Astrometry Code}

Finally, we ran the positions through a publicly-available astrometry 
code, ATPa (G. Anglada Escud\'e 2010, in preparation).
This code was developed for use in the Carnegie Astrometric Planet Search. 

We de-rotated the positions reported by {\it img2xym\_HRC} using the 
nominal orientation, so that the (x,y) axes correspond to ($-$RA, Dec)
respectively. We wrote out the coordinates in the correct format to
input into the ATPa astrometric solution tool. 
We used the nominal HRC plate scale. Since the geometric distortions
have already been removed, we used calibration rank 1 in ATPa, which
corrects for linear offsets, rotations, and magnification changes.

Because this code is still under development, and we ran it as a black box,
we use it only as a check that our other codes are producing reasonable 
results. Results are stated in Table~\ref{tbl-results} as model~T.
We do not include the ATPa results in our final parallax determination.

\subsection{Results}

The results of the realizations of these fitting codes are shown in
Table~\ref{tbl-results}. As described above, we ran many realizations
of each of these codes, using different reference stars for the transformation,
selectively excluding stars or entire visits, and allowing the plate scale and
rotation to be either fixed or free parameters.
We have chosen to present only the mean of the many realizations for each code.
Differences are minor. Individual estimates of distances range from 117 to
128~pc (the $\pm 1~\sigma$ ranges encompass 109 to 135~pc).

\begin{figure}
\plotone{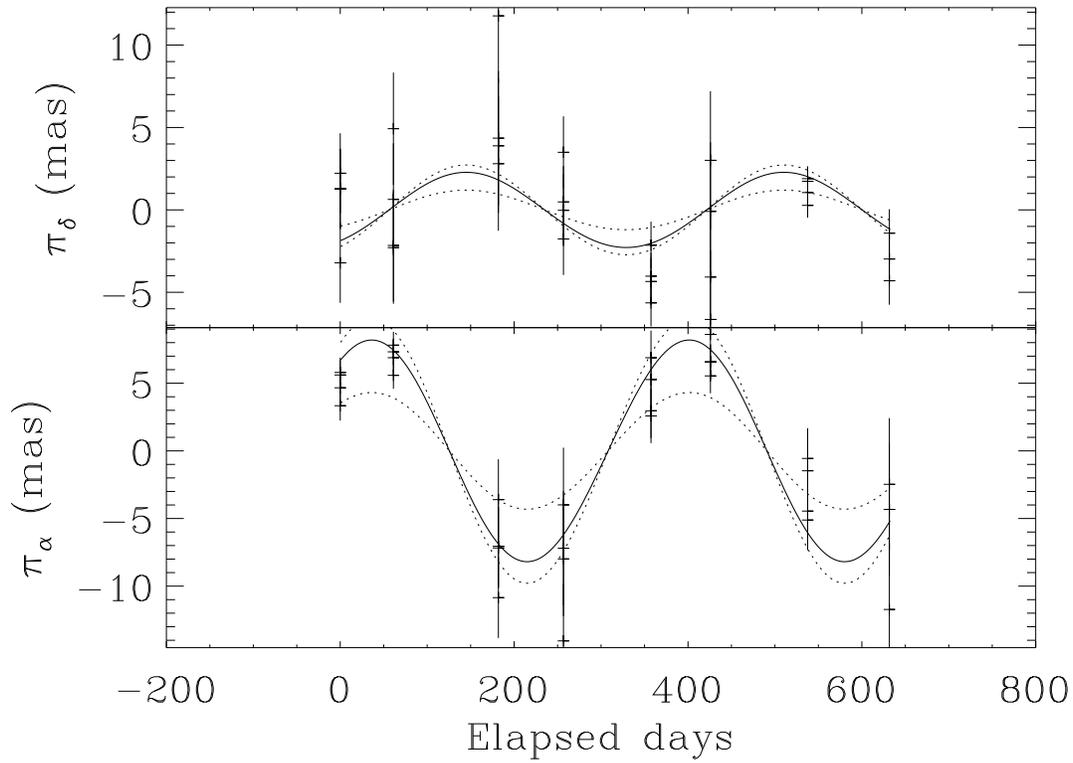}
\caption{Residuals to the motion of RX J1856 after subtracting the 
linear proper motion. The solid curves are the best fit parallactic ellipse;
the dotted curves are the 90\% confidence regions. The 4 observations at each
epoch are not averaged.
\label{fig-plx1}}
\end{figure}

\begin{figure}
\plotone{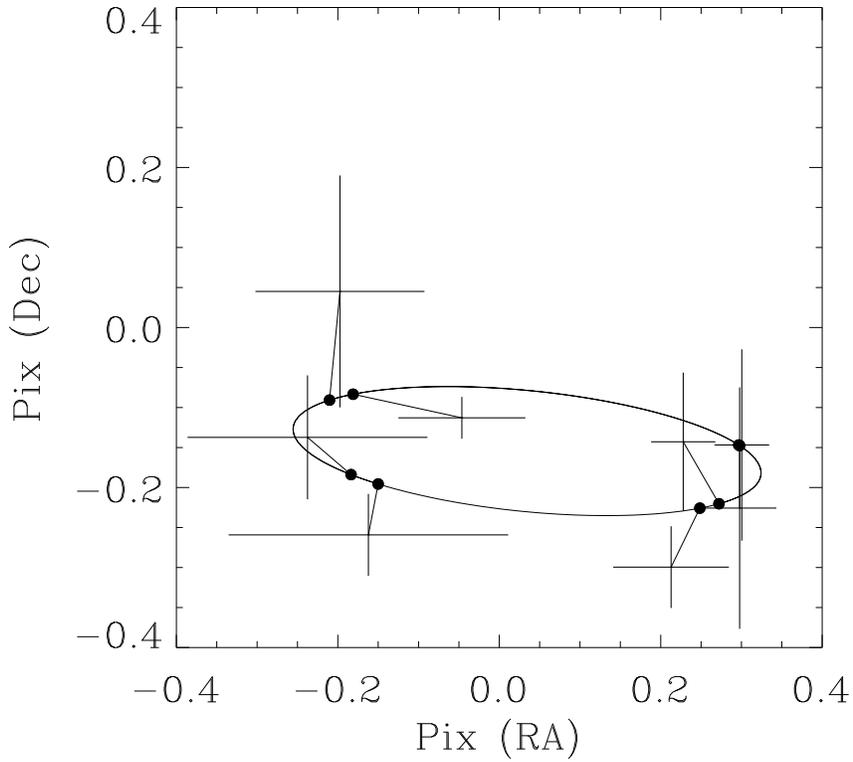}
\caption{Residuals to the motion of RX J1856 after subtracting the 
linear proper motion. The best fit parallactic ellipse is overplotted. 
The
thick lines represent the
mean residual positions at each of the 8 visits, after averaging the 
4 individual observations at each visit.
The dots are the predicted positions
along the parallactic ellipse at the time of observation.
\label{fig-plx3}}
\end{figure}

\begin{figure}
\plotone{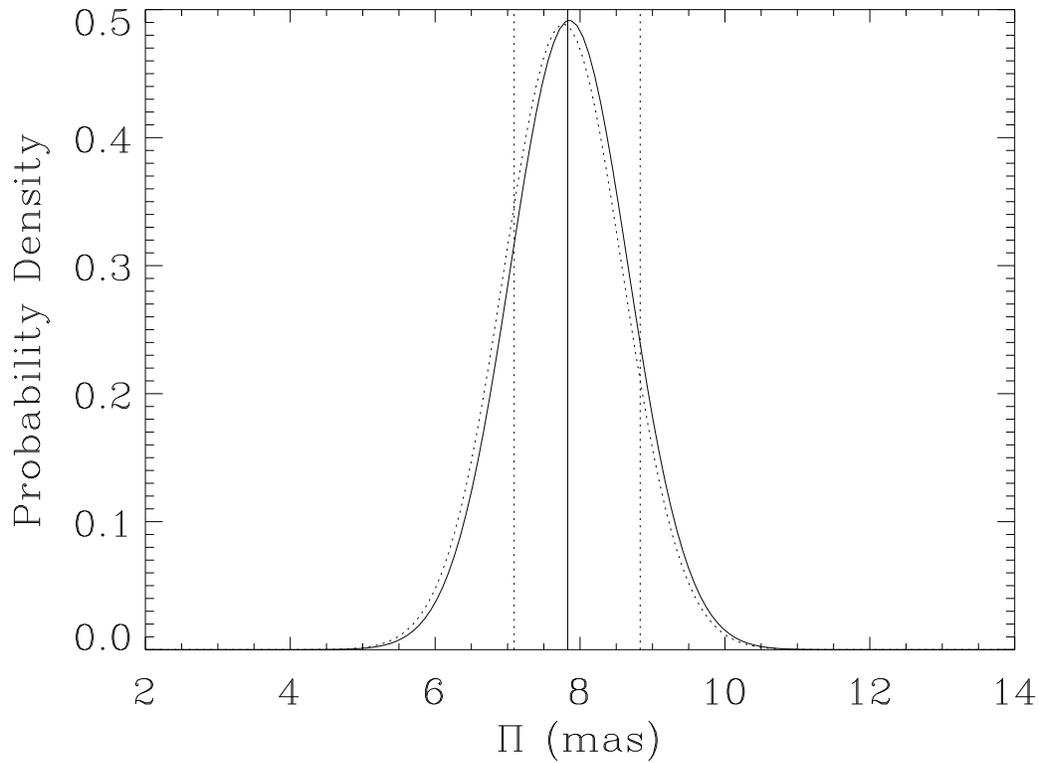}
\caption{Bayesian posterior probabilities and the estimate of parallax
of RX J185635$-$3754. The solid curve is the probability distribution
using flat priors; the dotted line is the probability distribution using
Jeffreys priors. 
The 68\% credible region centered on the highest probability
density for the flat prior solution is enclosed by the vertical
dotted lines for case i errors.
The probability density is per mas.
\label{fig-bpar}}
\end{figure}

\begin{figure}
\plotone{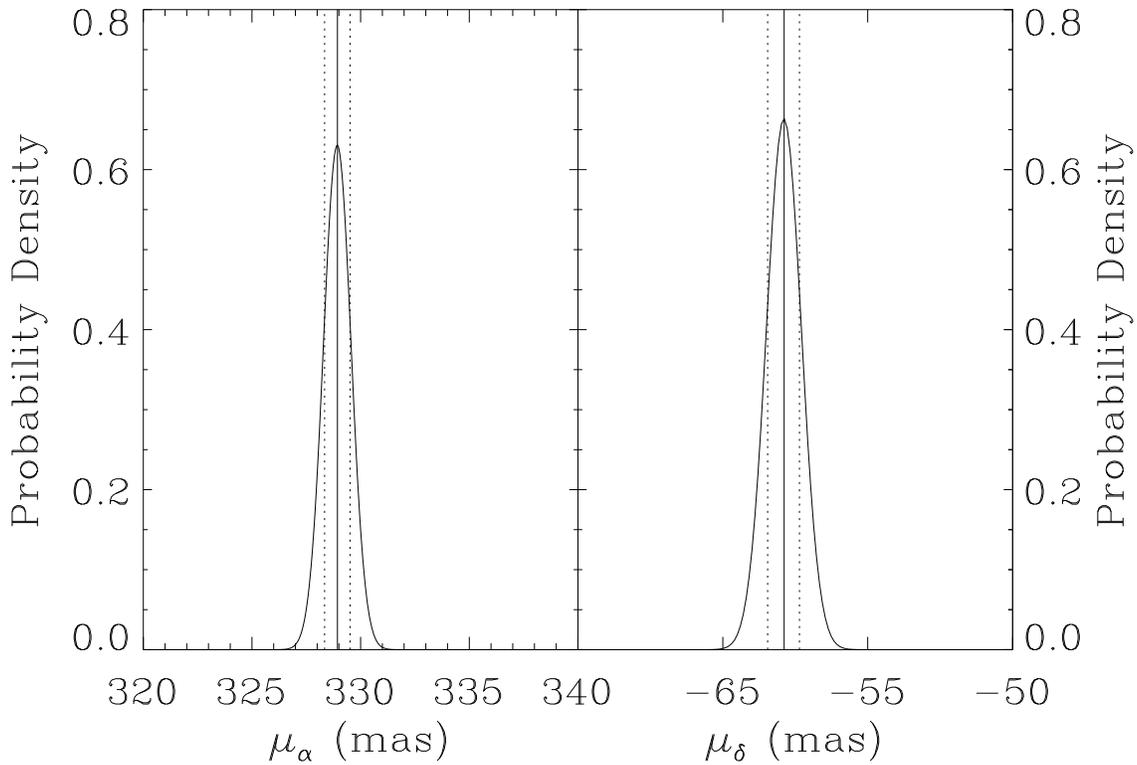}
\caption{Bayesian posterior probabilities and the estimate of the proper motion
in right ascension and declination
of RX J185635$-$3754 using the flat prior.
The 68\% credible region centered on the highest probability
density is enclosed by the vertical dotted lines for case i errors.
The probability density is per mas.
\label{fig-bmu}}
\end{figure}

The parallactic motion in RA and Dec are shown in 
Figure~\ref{fig-plx1}. The full 2-dimensional representation of the
parallactic ellipse is shown in Figure~\ref{fig-plx3}.
The Bayesian estimates for these parameters are shown in Figures~\ref{fig-bpar}
and \ref{fig-bmu}. The differences in the Bayesian estimates of the parallax
for the two choices of priors, 0.6\%, is negligble.  The choice of prior
for the parallax has ne effect on the proper motion estimate.
We use the mean of the two parallax estimates
in determing the weighted mean of all the models.

\begin{figure}
\plotone{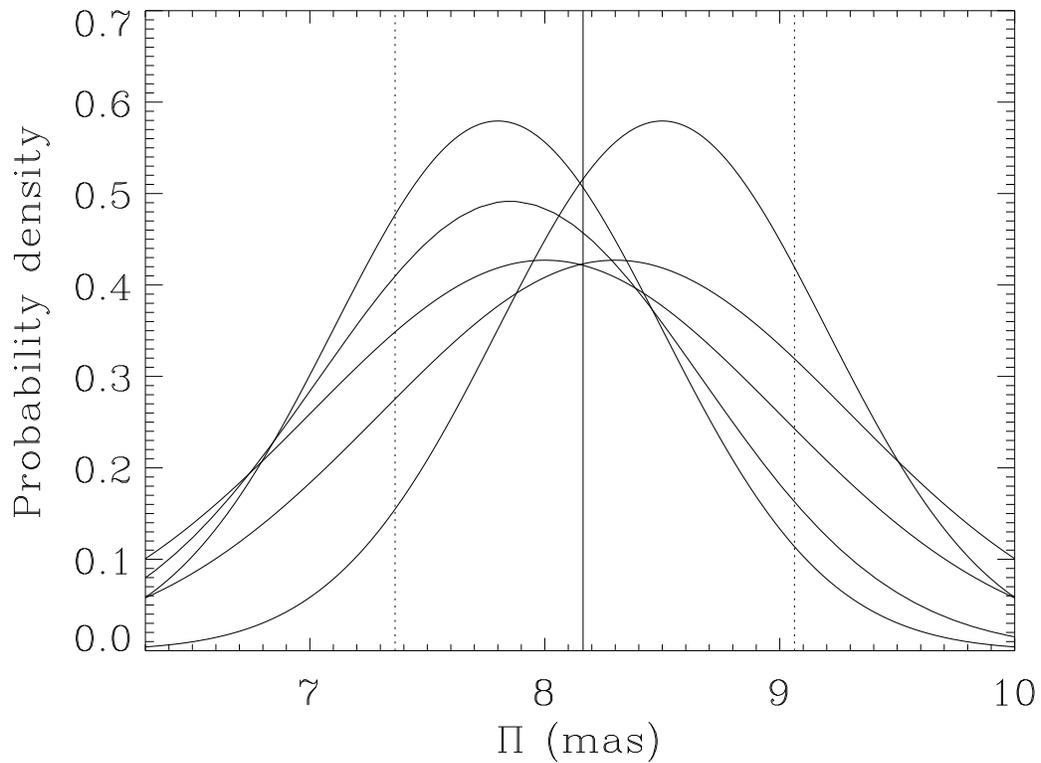}
\caption{Probability distribution functions for the parallax
of RX J185635$-$3754 from
the independent astrometric models.
The units are probability density per mas.
The thick vertical line is the weighted mean, 
which we adopt as the parallax. The dotted vertical lines encompass our
adopted $1~\sigma$ confidence region (see text).
\label{fig-plxpdf}}
\end{figure}

It is not clear which, if any, of the codes is to be preferred. 
The parallaxes of the A and B models all agree to within their $\pm$1~$\sigma$
confidence regions.
In the absence of a clear choice, we opt to view these as independent
measurements, and 
take a weighted mean for the final result. From this mean we exclude 
model T, which we used to corroborate our results.
The probability distributions of the independent models are shown in
Figure~\ref{fig-plxpdf}.

The five measurements of parallax agree to well within their 1~$\sigma$
uncertainties, with a standard deviation of 0.24~mas.
The weighted mean parallax of 8.18~mas corresponds to a distance of 123~pc.
Because the 5 measurements all start with the same data, they are not truly
independent measurements of data selected from a sample population. Therefore,
we adopt the mean error of the individual measurements, 0.8~mas, as
representative of the true uncertainty. The corresponding distance is
123\ud{11}{13}~pc.

Any corrections for systematic errors made by assuming the plate scale and 
orientation of the reference image are $<6\times 10^{-5}$ and $<\pm0.0055$
degrees. These are far smaller than our mesasurement errors, and are
inconequential.


Note, however, that this is a differential measurement, and that our
measured parallax is only a lower limit to the true parallax. Most if
not all of the reference objects are likely to be galactic stars.
\cite{KvKA02} have estimated the distance of the background stars using a
$B$ vs.\ $B-R$ color-magnitude diagram. They conclude that most of the 
objects are consistent with dwarfs at a distance of order 10~kpc, or K giant
members of the Sgr dwarf galaxy at a distance of 25~kpc. 
The six brightest of our reference objects have $B$ and $R$ magnitudes in the
NOMAD-1 catalog; these are consistent with the \cite{KvKA02} picture.
Star 9 lies near the 25~kpc giant branch; stars 6, 13, 17, and 21 lie near
the 10-25~kpc main sequence; star 12 lies about 2 mag below the 25~kpc main
sequence. We constructed a color-magnitude diagram using the 10 stars
in common between the HRC images and the F606W WFPC2 images \citep{W01}. 
We use the 3 of these that also
have $B$ and $R$ magnitudes for a crude calibration.
If stars 6 and 21 define the 10-25~kpc main sequence (see
above), the other 8 stars
are 2-3 mag fainter at similar colors. These could main sequence stars
at distances of 70-100~kpc (and a galactic scale height of $20-30$~kpc), 
or white dwarfs at distances of order 10~kpc.
We conclude from this
crude analysis that the mean absolute parallax of the reference stars is
$<$0.1~mas, and that the relative parallax is unlikely to be significantly
underestimated. Nonetheless, we will add 0.1~mas to the error budget on the
positive side. The consequence is to decrease the lower bound on the
distance by about 2~pc. Our best estimate of the distance is
123\ud{11}{15}~pc.

The $\pm 1~\sigma$ confidence region encompasses 108 to 134~pc. 
This is in good agreement with the previously published value of
117$\pm$12~pc, and with distances estimated from the interstellar extinction.
The proper motion (Table~\ref{tbl-results}) compares favorably to that
published by \citet{WL02}.

The F475W magnitude of the neutron star is 25.36$\pm$0.15. This magnitude
is the mean of the corrected instrumental magnitudes in each visit from the
{\it img2xym\_HRC} code, converted to apparent magnitude using the photometric
zero point for the F475W filter.
The uncertainty is the standard deviation of the measurements in the 8 visits.
There is no evidence for variability.

As a by-product of this analysis, we have measured
magnitudes, proper motions, and
parallaxes for the field stars. These are shown in Table~\ref{tbl-allstars}.
Five of the reference stars have proper motions that are
formally significant at the 
3~$\sigma$ level (one at $>5.0~\sigma$), with values ranging from 4 to
14~mas~yr$^{-1}$.
None of the reference stars has a measurable parallax
at the $>3~\sigma$ level. There is evidence for photometric variability of
stars 2 and 21.

\section{Discussion}

The proper motion and parallax are in excellent agreement with those published
earlier by \citet{WL02}. The proper motions agree at the 0.2\% level, and the
parallaxes agree to better than 6\%, both well within the formal uncertainties.
These two measurements were made using different instruments.
Their excellent agreement 
certainly speaks to the great care that has gone into the
astrometric calibration of both the HRC \citep{AK04} and the WFPC2
\citep{H95}, as well as the astrometric stability of the HST as a whole. 

That the new HRC results are not superior to the determination from the WFPC2,
despite the 
smaller pixel size of in the HRC, can be attributed to two factors:
the smaller plate scale
is offset by the larger number of reference stars in the larger WFPC2 field,
and the proper motions are better-determined in the WFPC2 images bcause of the
longer 4.5 year baseline.
Combining these two independent
measures results a weighted mean parallax of 8.33$\pm$0.6~mas
(120\ud{8}{10}~pc). A combined analysis of the two data sets is likely to
yield vastly reduced errors because the baseline for the proper motion will more
than double to 8.5 years. Better-determined proper motions will yield a
more accurate transformation, and ultimately a better measure of the parallax.

Obtaining the parallax of this neutron star is only a means to an end.
With a distance
known to 7\%, the largest uncertainty in the radius of this
neutron star lies in the atmospheric or surface emission models.
In the context of simple two blackbody
models the radius R$_\infty$ is 16.8\ud{1.1}{1.4}~km, where this uncertainty 
reflcts the statistical uncertainty in the distance
(as discussed by \cite{P01}, the
radius depends critically on the ill-defined temperature of the cool component).
The true uncertainty is dominated by systematics in the models. 
This is still consistent with many equations of state, but combined with
information for X-ray bursts and quiescent low-mass X-ray binaries, it can be
used to significantly constrain the nuclear equation of state 
\citep{SLB10}.



\acknowledgments

This investigation was based on observations made with the NASA/ESA
Hubble Space Telescope, obtained from the MAST data archive at the
Space Telescope Institute. STScI is operated by the
Association of Universities for Research in Astronomy, Inc. under 
NASA contract NAS 5-26555.
We thank Jay Anderson for providing the $img2xym\_HRC$ software.
FMW thanks Guillem Anglada Escud\'e for providing the ATPa code, explaining
its operation, and for useful comments on the text.
A statistical reviewer suggsted the use of the Jeffreys prior.
FMW is grateful to the Sonderforschungsbereich
(SFB/TR 7 on Gravitational Wave Astronomy), 
which is funded by the German Research Foundation (DFG),
for support during a visit to Jena, where much of this work was completed.
VVH, TE, and RN would like to thank DFG for
support in SFB/TR 7 on Gravitational Wave Astronomy.

\begin{figure}
\plotone{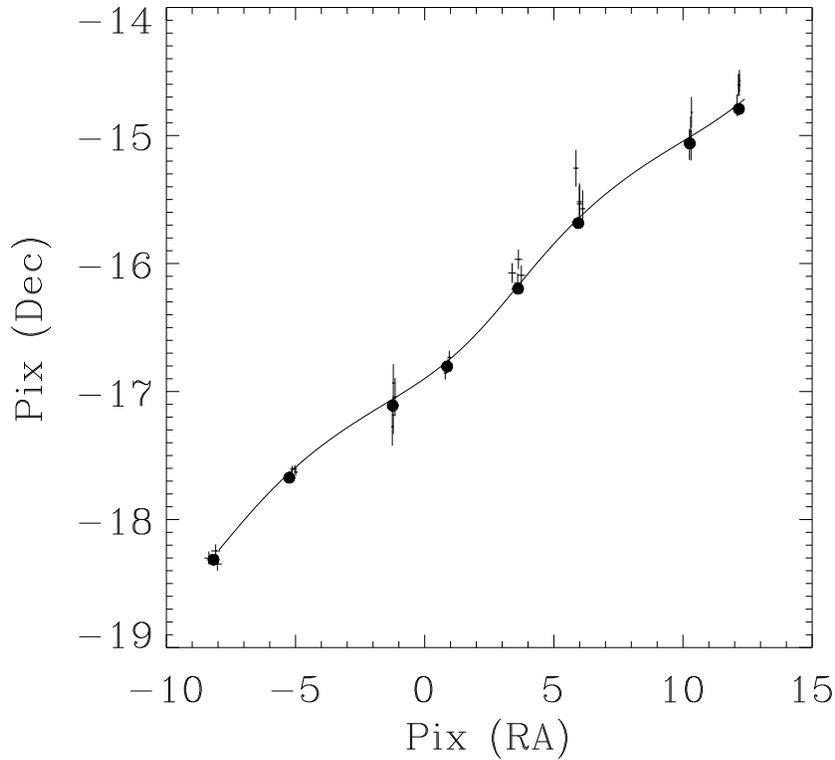}
\caption{Motions of RX J1856 after transforming the 32 images to
the same reference. The solid curve is the best fit parallax + proper motion;
the proper motion and parallactic wobble are clearly evident. The motion is
towards the lower left.
The dots show the predicted positions at the times of observation.
The units are in corrected pixels, 28.27 milli-arcsec.
Note that the declination scale is 5 times that of the RA scale.
\label{fig-pm}}
\end{figure}

\clearpage

\clearpage

\clearpage


\clearpage
\begin{deluxetable}{ll}
\tablewidth{0pt}
\tablecaption{Epochs of HRC Exposures. \label{tbl-times}}
\tablehead{\colhead{Start (UT)} &\colhead{Start (MJD)}
}
\startdata
2002-09-01 04:48:45 & 52518.201 \\
2002-11-01 08:13:59 & 52579.343 \\ 
2003-03-02 02:50:49 & 52700.119 \\ 
2003-05-15 22:54:12 & 52774.954 \\ 
2003-08-24 16:00:10 & 52875.667 \\ 
2003-11-01 00:21:29 & 52944.015 \\ 
2004-02-20 21:06:06 & 53055.879 \\ 
2004-05-24 19:17:10 & 53149.804  \\
\enddata
\end{deluxetable}

\clearpage

\begin{deluxetable}{rrrrrrr}
\tablewidth{0pt}
\tablecaption{Recovery of Simulated Data. \label{tbl-sims}}
\tablehead{\multicolumn{3}{c}{Input values} & \multicolumn{3}{c}{Output values} & \colhead{N\tablenotemark{a}}\\
\colhead{$\mu_x$} &\colhead{$\mu_y$} &\colhead{$\Pi$} & 
\colhead{$\mu_x$} &\colhead{$\mu_y$} &\colhead{$\Pi$}\\
\colhead{mas/yr} &\colhead{mas/yr} &\colhead{arcsec} & 
\colhead{mas/yr} &\colhead{mas/yr} &\colhead{arcsec}
}
\startdata
-307.2 & 51.2 & 0.0080 & -307.3 $\pm$ 1.2 & 51.5 $\pm$ 1.3 & 0.0079 $\pm$ 0.0009 & 2204\\

-309.8 & 51.6 & 0.0060 & -309.7 $\pm$ 1.2 & 52.0 $\pm$ 1.3 & 0.0069 $\pm$ 0.0010 & 2001\\
-309.8 & 51.6 & 0.0060 & -309.7 $\pm$ 1.2 & 52.0 $\pm$ 1.3 & 0.0060 $\pm$ 0.0010 & 2001\\
-309.8 & 51.6 & 0.0050 & -309.7 $\pm$ 1.2 & 52.0 $\pm$ 1.3 & 0.0050 $\pm$ 0.0010 & 2001\\
-309.8 & 51.6 & 0.0040 & -309.7 $\pm$ 1.2 & 52.1 $\pm$ 1.3 & 0.0040 $\pm$ 0.0010 & 2001\\
-309.8 & 51.6 & 0.0030 & -309.6 $\pm$ 1.2 & 52.0 $\pm$ 1.3 & 0.0031 $\pm$ 0.0009 & 2001\\
-309.8 & 51.6 & 0.0020 & -309.5 $\pm$ 1.2 & 52.0 $\pm$ 1.4 & 0.0022 $\pm$ 0.0009 & 2001\\
-309.8 & 51.6 & 0.0010 & -309.5 $\pm$ 1.2 & 52.1 $\pm$ 1.3 & 0.0011 $\pm$ 0.0009 & 2001\\

\enddata
\tablenotetext{a}{Number of trials run.}
\end{deluxetable}

\clearpage

\begin{deluxetable}{llrll}
\tablewidth{0pt}
\tablecaption{Proper Motion and Parallax Measurements. \label{tbl-results}}
\tablehead{
\colhead{Model} & \colhead{$\mu$} & \colhead{PA} & \colhead{$\Pi$} & \colhead{Notes\tablenotemark{a}}\\
\colhead{}  & \colhead{(mas/yr)} & \colhead{(deg)} & \colhead{(mas)} & \colhead{}}
\startdata
A1  & 332.4 $\pm$0.7 & 100.8 $\pm$ 0.1 & 8.5 $\pm$0.7 & \\  
A2  & 329.5 $\pm$1.0 & 100.3 $\pm$ 0.2 & 8.0 $\pm$1.0\\  

A3  & 329.3 $\pm$1.0 & 100.3 $\pm$ 0.2 & 8.3 $\pm$1.0\\ 
A4  & 330.7 $\pm$1.0 & 100.1 $\pm$ 0.2 & 8.2 $\pm$0.6\\  
B   & 333.9 $\pm$0.7 & 100.0 $\pm$ 0.1 & 7.8 $\pm$0.8 & \\
T   & 332.6 $\pm$0.1 & 100.8 $\pm$ 0.1 & 7.8 $\pm$0.4 & not included in mean\\
mean& 331.2 $\pm$2.0 & 100.3 $\pm$ 0.3 & 8.16 $\pm$0.27 \\
 \\
P1 & 332.3 $\pm$0.4 & \multicolumn{1}{l}{100.5} & 8.5 $\pm$0.9& \citet{WL02}\\

\enddata

\end{deluxetable}

\clearpage


\begin{deluxetable}{rrrrrrrl}
\tablewidth{0pt}
\tablecaption{Physical Properties of the Field Stars. \label{tbl-allstars}}
\tablehead{
\colhead{Star} & \colhead{RA\tablenotemark{a}} & \colhead{Dec} & \colhead{mag} & \colhead{$\mu_\alpha$} & \colhead{$\mu_\delta$} & \colhead{$\Pi$} & ID\tablenotemark{b}  \\
\colhead{} & \multicolumn{2}{c}{(J2000)} &  & \colhead{mas/yr} & \colhead{mas/yr} & \colhead{mas} & 
}
\startdata
 0 & 18 56 36.531 & -37 54 20.19 & 20.72$\pm$0.02 &  1.0$\pm$0.7 & -0.0$\pm$1.1 &  0.6$\pm$0.2\\
 1 & 18 56 36.612 & -37 54 30.47 & 24.41$\pm$0.05 &  9.3$\pm$1.9 &  0.9$\pm$0.9 &  0.0$\pm$0.6\\
 2 & 18 56 36.599 & -37 54 30.41 & 22.89$\pm$0.39 &  1.8$\pm$1.7 & -4.0$\pm$0.8 &  0.4$\pm$0.5\\
 3 & 18 56 36.465 & -37 54 22.92 & 24.35$\pm$0.09 &  3.2$\pm$2.3 &  0.2$\pm$1.5 &  0.2$\pm$0.4 & W: 112\\
 4 & 18 56 36.215 & -37 54 32.19 & 24.52$\pm$0.09 &  0.5$\pm$2.6 &  6.6$\pm$1.1 & -0.2$\pm$0.4 & W: 113\\
 5 & 18 56 36.026 & -37 54 29.45 & 24.74$\pm$0.08 &  3.2$\pm$1.9 &  4.6$\pm$1.1 & -0.7$\pm$0.4 & W: 114\\
 6 & 18 56 36.052 & -37 54 31.75 & 21.23$\pm$0.13 &-11.6$\pm$2.3 &  7.1$\pm$0.9 &  0.8$\pm$0.3 & C: 24\\
 7 & 18 56 36.202 & -37 54 46.64 & 24.41$\pm$0.13 &  4.3$\pm$2.2 & -4.4$\pm$1.2 &  0.1$\pm$0.3\\
 8 & 18 56 36.139 & -37 54 45.02 & 25.75$\pm$0.02 &  8.9$\pm$6.0 & -0.4$\pm$6.2 & -2.0$\pm$2.4\\
 9 & 18 56 35.835 & -37 54 24.01 & 17.69$\pm$0.09 & -3.3$\pm$2.2 & -1.0$\pm$1.2 & -0.1$\pm$0.2 & WM: L\\ 
10 & 18 56 35.653 & -37 54 23.62 & 22.69$\pm$0.10 & -4.2$\pm$1.6 &  1.4$\pm$1.4 &  0.0$\pm$0.2\ & W: 106\\ 
12 & 18 56 35.408 & -37 54 27.76 & 21.31$\pm$0.09 &  3.0$\pm$1.3 & -0.2$\pm$0.4 &  0.1$\pm$0.1 & WM: J\\ 
13 & 18 56 35.537 & -37 54 48.73 & 18.12$\pm$0.03 &  5.4$\pm$1.1 &  0.4$\pm$0.6 &  0.3$\pm$0.3 & WM: C\\ 
14 & 18 56 35.286 & -37 54 34.48 & 24.64$\pm$0.10 &  9.3$\pm$2.0 & -3.2$\pm$1.3 &  0.1$\pm$0.4 & C: 23\\ 
15 & 18 56 35.366 & -37 54 46.23 & 25.68$\pm$0.06 &  2.2$\pm$2.4 &  1.5$\pm$5.8 &  1.5$\pm$1.1 & W: 128\\
16 & 18 56 34.976 & -37 54 27.96 & 21.76$\pm$0.11 &  2.6$\pm$1.0 & -1.9$\pm$0.7 &  0.1$\pm$0.2 & W: 28\\ 
17 & 18 56 35.221 & -37 54 46.66 & 18.79$\pm$0.09 & -3.6$\pm$0.6 &  1.4$\pm$0.7 &  0.6$\pm$0.2 & WM: D\\ 
18 & 18 56 35.268 & -37 54 50.44 & 21.62$\pm$0.03 &  4.4$\pm$1.5 & -1.5$\pm$1.1 & -0.1$\pm$0.2\\
19 & 18 56 34.900 & -37 54 35.50 & 21.85$\pm$0.10 &  0.7$\pm$0.6 & -0.7$\pm$0.9 &  0.6$\pm$0.2\\
20 & 18 56 35.056 & -37 54 24.23 & 25.47$\pm$0.16 &  6.3$\pm$2.1 &  2.5$\pm$2.8 & -0.7$\pm$0.4\\
21 & 18 56 34.899 & -37 54 24.11 & 19.25$\pm$0.18 &  2.1$\pm$2.7 & -1.4$\pm$1.0 &  0.7$\pm$0.3 & WM: I \\
 & \\
NS & 18 56 35.795 & -37 54 35.54 & 25.33$\pm$0.15 &  8.2$\pm$0.2 &326.6$\pm$0.5& -61.9$\pm$0.4 \\
\enddata
\tablenotetext{a}{Positions are referenced to the NOMAD-1 catalog position of star 9, and are epoch 2003.624.} 
\tablenotetext{b}{Previous identifications.  C: \cite{CMS97}; W: \cite{W01}; WM: \cite{WM97}}
\end{deluxetable}

\end{document}